\documentclass[twocolumn,superscriptaddress,nofootinbib,aps,floatfix,amsfonts,preprintnumbers,amsmath,amssymb,groupedaddress]{revtex4} 
\usepackage{graphicx, amsthm, multirow} 
\usepackage{url}
\usepackage{color}
\usepackage{dsfont}
\usepackage{stackrel}
\usepackage{bbm}
\usepackage{fancybox}
\usepackage{mathtools}
\usepackage{mdframed}
\usepackage{physics}
\usepackage{float}
\usepackage{bbold}
\usepackage{lpic}
\usepackage{array}
\usepackage{hyperref}

\definecolor{light-gray}{gray}{0.65}

\newcolumntype{L}[1]{>{\raggedright\let\newline\\\arraybackslash\hspace{0pt}}m{#1}}

\def\permille{\ensuremath{{}^\text{o}\mkern-5mu/\mkern-3mu_\text{oo}}}

\begin{document}

\title{\vspace*{1cm} Comment on Daya Bay's definition and use of $\Delta m^2_{ee} $}


\author{Stephen J. Parke}
\email{parke@fnal.gov orcid:0000-0003-2028-6782}
\affiliation{Theoretical Physics Department, Fermi National Accelerator Laboratory, P.O.Box 500, Batavia, IL 60510, USA}

\author{Renata Zukanovich Funchal}
\email{zukanov@if.usp.br orcid:0000-0001-6749-0022}
\affiliation{Instituto de F\'{\i}sica, Universidade de S\~ao
  Paulo, C.\ P.\ 66.318, 05315-970 S\~ao Paulo, Brazil }

\begin{abstract} 
We comment on Daya Bay's latest definition of the effective $\Delta m^2$ for short baseline reactor  $\bar{\nu}_e$ disappearance experiments used in \cite{Adey:2018zwh}.
\end{abstract}
\preprint{FERMILAB-Pub-19-078-T} 
\date{\today}
\maketitle

 \hspace*{0.5cm} 
 In \cite{Adey:2018zwh}, Daya Bay (DB) uses their latest definition of $\Delta m^2_{ee}$ which obfuscates the simple relationship between such an effective $\Delta m^2$ and the fundamental parameters of the neutrino sector. Furthermore, this definition of $\Delta m^2_{ee}$ is  baseline divided by neutrino energy  ($L/E$) dependent. Dependence on L/E implies dependence on the proper time between production and detection of the observed neutrinos, i.e. the proper age  of the neutrinos \cite{propertime}. This new definition is approximately constant for the Daya Bay experiment ($L/E <  1$ km/MeV), unlike DB's earlier definition,  \cite{DB1}. However, for  the JUNO experiment ($6< L/E < 25$ km/MeV),  currently under construction,  this new definition has a $\sim$ 1\%  jump between smallest and largest  $L/E$ values for the observed neutrinos. The expected precision on the measurement of  $\Delta m^2_{ee}$  at JUNO is 0.5\%, see \cite{An:2015jdp},  and therefore comparable in size to this jump.
 

Daya Bay's latest definition
of $\Delta m^2_{ee}$ is given by
\begin{eqnarray}
\Delta m^2_{ee} (\rm DB2)  & \equiv & \nonumber \\
& & \hspace{-2cm}  \Delta m^2_{32} + \frac{2E}{L} \arctan{\left( \frac{\sin2\Delta_{21}}{\cos 2 \Delta_{21}+\tan^2 \theta_{12}}\right) }  \label{eq:db2}
\end{eqnarray}
using $\Delta_{jk} \equiv \Delta m^2_{jk}L/(4E)$, see \cite{An:2015rpe}.  $\Delta m^2_{ee} (\rm DB2) $ is clearly $L/E$ (proper time of the neutrino)  dependent  {\it and} it is far from transparent the relationship to the fundamental neutrino parameters.

 The original definition of an effective $\Delta m^2$ for $\nu_e$ disappearance experiments, $\Delta m^2_{ee}$, given by Nunokawa, Parke and Zukanovich Funchal (NPZ) , is simply \cite{Nunokawa:2005nx}
 \begin{align}
& \Delta m^2_{ee} (\rm NPZ)   \equiv  \cos^2 \theta_{12} \Delta m^2_{31} + \sin^2 \theta_{12} \Delta m^2_{32} \, , \label{eq:npz}  \\[1mm]
  & \quad =    \Delta m^2_{31} - \sin^2 \theta_{12}  \Delta m^2_{21}  =   \Delta m^2_{32} +\cos^2 \theta_{12}  \Delta m^2_{21} \, . \nonumber 
\end{align}
The $L/E$ {\it independence} of this definition is  manifest and so is the relationship to the fundamental parameters of the neutrino sector.   $ \Delta m^2_{ee} (\rm NPZ) $ is ``the $\nu_e$ average of $\Delta m^2_{31}$ and $\Delta m^2_{32}$''.  RENO uses this definition   \cite{RENO:2015ksa}.



 
Since there is no review of $\Delta m^2_{ee}$ in the PDG, clarification of the relationship between these different definitions of $\Delta m^2_{ee}$ is pertinent  for understanding short baseline reactor neutrino oscillation physics.  To start, consider the small and large $L/E$ limits of $\Delta m^2_{ee} (\rm DB2)$:
 for  $\frac{L}{E} \ll \frac{2\pi}{\Delta m^2_{21}} \approx 15 ~{\rm km/MeV}$ ($ \Delta_{21} \ll 1$),  
\begin{eqnarray}
\Delta m^2_{ee} (\rm DB2) & = & \Delta m^2_{ee} (\rm NPZ) \label{eq:identical} \\[1mm]
&& \hspace{-2cm} +  \Delta m^2_{21}\, (\cos 2 \theta_{12} \sin^2 2\theta_{12} /6) \,\Delta^2_{21}
 + ... , \nonumber  \\
 & & \hspace{-3.cm}  {\rm whereas ~for} ~ \frac{L}{E} \ge \frac{2\pi}{\Delta m^2_{21}} \approx 15 ~{\rm km/MeV} \quad {\rm one ~finds}  \nonumber \\
 \Delta m^2_{ee} (\rm DB2) & \approx &  \Delta m^2_{31}\,.
\end{eqnarray}


\begin{figure}[b]
\vspace*{-0.5cm}
\begin{center}
\includegraphics[width=0.36\textwidth]{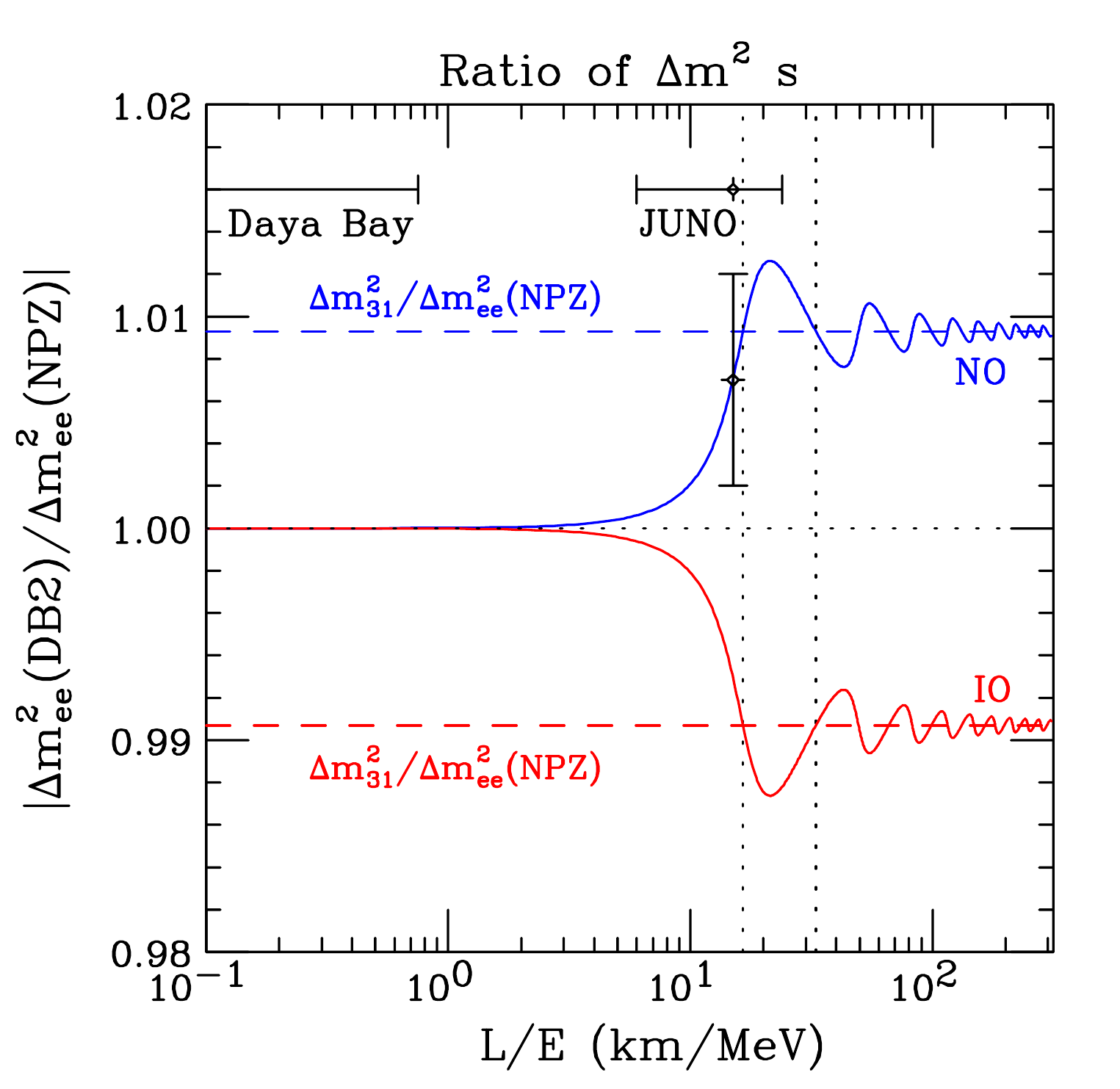}
\end{center}
\caption{The ratio of $\Delta m^2_{ee} (\rm DB2)$ to $\Delta m^2_{ee} (\rm NPZ)$ verse $L/E$ for the normal ordering (NO) and the inverted ordering (IO).  The horizontal black lines, labelled Daya Bay and JUNO, show the $L/E$ ranges of these experiments.   The vertical black line, labelled JUNO, represents the fractional uncertainty expected from the JUNO experiment (the vertical position was arbitrarily chosen for illustration purposes only). The vertical dotted lines are at $\Delta_{21}=\pi/2, ~\pi$, from left to right. 
}
\label{fig:ratio}
\end{figure}

In Fig. 1, the ratio of $\Delta m^2_{ee} (\rm DB2)$ to $\Delta m^2_{ee} (\rm NPZ)$ is plotted as a function of $L/E$.  Clearly, for the Daya Bay experiment, $L/E < 1$ km/MeV, these two definitions are essentially identical as the second term in eqn.  \ref{eq:identical} is always smaller  than 1\permille  ~ of $\Delta m^2_{21}$.  Whereas for the JUNO experiment, $6 < L/E <  25 $ km/MeV, there is a significant jump, $\sim$1\%, in the value $\Delta m^2_{ee} (\rm DB2)$ between the smallest and largest $L/E$.   

In summary, Daya Bay's new definition of $\Delta m^2_{ee}$, eq. \ref{eq:db2}, does not manifestly show  the simple relationship to the fundamental parameters of the neutrino sector for short baseline reactor experiments, such as Daya Bay and RENO. Nor is it useful for future medium baseline experiments like JUNO due to the $\sim$1\% jump, precisely in the  $L/E$ range of this experiment \cite{uniqueness}. The original definition of  $\Delta m^2_{ee}$, eq. \ref{eq:npz}, is clearly $L/E$ independent, the relationship to the fundamental parameters of the neutrinos sector is manifest and useful for both the short, Daya Bay and RENO, and medium baseline, JUNO, reactor neutrino experiments.

\begin{acknowledgments}
We acknowledge discussions with our long time collaborator Hiroshi Nunokawa who is currently a member of the JUNO collaboration.
Fermilab is operated by the Fermi Research Alliance under contract no. DE-AC02-07CH11359 with the U.S. Department of Energy. 
 SP thanks IFT of Madrid for wonderful hospitality while this comment was written..
This project has received funding/support from the European Union’s Horizon 2020
research and innovation programme under the Marie Sklodowska-Curie grant agreement
No 690575 and. No 674896.
RZF was partially supported by Funda\c{c}\~ao de Amparo \`a
Pesquisa do Estado de S\~ao Paulo (FAPESP) and Conselho Nacional de
Ci\^encia e Tecnologia (CNPq). 
\end{acknowledgments}


\begin{thebibliography}{99}
\bibitem{propertime}
  L/E is proportional to the proper time of the neutrino, as $\tau_\nu = 0.3  \left(\frac{L/E}{1~\rm km/MeV}\right) \left( \frac{m_\nu}{\, 100 \, \rm meV} \right) \times 10^{-12}$ sec.
\bibitem{uniqueness}
Until JUNO determines the mass ordering, which is highly non-trivial due to stringent requirements  on the resolution and linearity  of the neutrino  energy reconstruction, $\Delta m^2_{ee}$ is the only atmospheric $\Delta m^2$ that JUNO can report without having to give separate  measurements for each mass ordering, as would be needed for $\Delta m^2_{31}$ (or $\Delta m^2_{32}$), since $\Delta m^2_{31}= \pm | \Delta m^2_{ee}| +\sin^2 \theta_{21} \Delta m^2_{21}$.
\end{thebibliography}
\end{document}